\documentclass[12pt,twoside,fleqn]{article}
\input BoxedEPS
\SetRokickiEPSFSpecial  
\HideDisplacementBoxes

\newcommand{\AmS}{{\protect\the\textfont2
  A\kern-.1667em\lower.5ex\hbox{M}\kern-.125emS}}
\newcommand{\half}{{\textstyle{1\over2}}}
\newcommand{\beq}{\begin{equation}}
\newcommand{\eeq}{\end{equation}}

\def\beq{\begin{equation}}
\def\eeq{\end{equation}}
\def\bi{\begin{itemize}}
\def\ei{\end{itemize}}
\def\beqar{\begin{eqnarray}}
\def\eeqar{\end{eqnarray}}

\newcommand{\Hh}{{\cal H\;}}

\newcommand{\Pp}{{\cal P\;}}
\newcommand{\bcP}{\mbox{\boldmath$\cal P$}}

\newcommand{\bW}{\mbox{\boldmath$\omega$}}
\renewcommand{\r}{{\bf r}}
\newcommand{\R}{{\bf R}}
\newcommand{\B}{{\bf B}}

\newcommand{\f}{{\bf f}}

\newcommand{\bu}{{\bf u}}
\newcommand{\bv}{{\bf v}}

\newcommand{\G}{{\bf G}}
\newcommand{\q}{{\bf q}}
\newcommand{\bP}{{\bf P}}

\newcommand{\bj}{{\bf j}}

\newcommand{\drh}{\dot{\rho}}

\newcommand{\rmd}{{\rm d\null}}

\def\boldnab{\mbox{\boldmath$\nabla$}}

\def\boldphi{\mbox{\boldmath$\phi$}}

\def\lra{\mathop{\hbox to .4in{\rightarrowfill}}}

\title{Fluid Dynamical Profiles and Constants of Motion from D-Branes}
\author{R. Jackiw\\\small\it Center for Theoretical Physics\\\small\it 
Massachusetts Institute of Technology\\\small\it  Cambridge, MA
~02139--4307%
\thanks{This work is supported in part by funds
provided by  the U.S.~Department of Energy (D.O.E.) under contract
\#DE-FC02-94ER40818.\qquad MIT-CTP \#2926}}

\begin{document}
\date{}
\maketitle

\thispagestyle{empty}

\begin{abstract}\noindent
Various fluid mechanical systems, governed by nonlinear differential
equations, enjoy a hidden, higher-dimensional dynamical Poincar\'e
symmetry, which arises owing to their descent from a Nambu-Goto
action.  Also, for the same reason, there are equivalence transformations
between different models.  These interconnections  are discussed in this
lecture, and are summarized in Fig.~\ref{fig:3} below.
\end{abstract}

\noindent
Having attended a few years ago the dedication of the Bogolyubov Institute
for Theoretical Physics, I am happy to be here again to commemorate the
90th anniversary of this emminent Kievan mathematician/physicist. These
days, fueled by the string program, there is a vigorous interchange between
mathematics and physics, which Bogolyubov would have been happy to
witness. His activities encompassed field theory, nonlinear systems, and
kinetic theory. Just these subjects have become linked in my work, which I
shall describe in this lecture, dedicated to his memory. 

I shall speak about several nonlinear equations of mathematical physics in
arbitrary spatial dimensions, which possess remarkable hidden symmetries and
unexpected constants of motion that allow construction of solutions, even
complete integration.  Moreover, the equations enjoy relationships with each
other that provide mappings of one onto another.  In the language currently
in use, there are ``dualities'' that relate completely different models. 
Finally, I shall describe how all these properties derive from an
``Ur-formulation'' in terms of extended objects -- d-branes -- in a higher
dimension.  The nonlinear equations that I shall discuss are pre-existing
and well known -- they are \emph{not} constructed to illustrate the
theory.  The observations I make about them are mostly new, though
some results in low spatial dimensionality were known previously.  The
work I describe was performed in collaboration with Bazeia~\cite{ref:1} and
Polychronakos~\cite{ref:2}; it is based on initial observations in this area
by Bordermann and Hoppe~\cite{ref:3}, as well as by Jevicki~\cite{ref:4}.

\section{Nonrelativistic Model}

The first equation that I shall consider describes nonrelativistic fluid
motion in $d$-spatial dimensions, $(d,1)$ space-time.  The matter density of
the fluid is $\rho(t,\r)$; its local velocity is $\bv (t,\r)$ and the current
$\bj (t, \r) = \bv (t, \r) \rho(t,\r)$ is linked to the density by a
continuity equation
\beq
\drh + \boldnab \cdot (\bv  \rho) = 0
\label{eq:1.1}
\eeq
(Over-dot denotes differentiation with respect to time.)  The velocity
satisfies the Euler equation, which relates the material time derivative of
$\bv$ to a force (per unit volume), that is, to the gradient of the
pressure $P$ divided by $\rho$
\beq
\dot{\bv} + \bv \cdot \boldnab \bv = \f = -\frac{1}{\rho} \boldnab P
\label{eq:1.2}
\eeq

For the nonrelativistic application, I shall be interested in a very special
force-law $\f$: its properties are, first, it arises from a pressure $P$ that
is a function only of $\rho$ (this corresponds to isentropic flow); second,
$P$ is of the polytropic form ($P \propto$ power of $\rho$), and finally,
third, the specific power law is the inverse power; that is, 
\beq
P(\rho) = -2 \lambda/m\rho
\label{eq:1.3}
\eeq
This is called the ``Chaplygin gas'' and corresponds to a sound speed
${\sqrt{2\lambda/m}}\,/{\rho}$ (hence we take $\lambda \ge 0$) and enthalpy
$\lambda/m\rho^2$ ($m$ is the mass)~\cite{ref:5}.

The Euler equation for the Chaplygin gas reads
\beq
\dot{\bv} + \bv \cdot \boldnab \bv =  -\frac{2 \lambda}{m} \frac{\boldnab
\rho}{\rho^3}
\label{eq:1.4}
\eeq
It is consistent to look for solutions without vorticity $(\boldnab \times
\bv=0$), so we write
\beq
\bv = \boldnab \theta/m
\label{eq:1.5}
\eeq
and replace (\ref{eq:1.4}) by Bernoulli's equation for the velocity
potential $\theta$, with a source term given by the enthalpy.
\beq
\dot{\theta} + \frac{(\boldnab \theta)^2}{2m} = \frac{\lambda}{\rho^2}
\label{eq:1.6}
\eeq
The gradient of (\ref{eq:1.6}) reproduces (\ref{eq:1.4}).

In summary, we are studying the irrotational and isentropic motion of the
Chaplygin gas.

Eqs.~(\ref{eq:1.1}) and (\ref{eq:1.6}) possess an action formulation with a
first-order (in time) Lagrangian.
\beq
L_{\rm NR} = \int \rmd r \left[ \theta \drh - \left(\rho \frac{(\boldnab
\theta)^2}{2m} + \frac{\lambda}{\rho}\right) \right]
\label{eq:1.7}
\eeq
Evidently the Hamiltonian is
\beq
H_{\rm NR} = \int \rmd r\,  {\cal H}_{\rm NR} = \int \rmd r \left(\rho \frac{(\boldnab
\theta)^2}{2m} + \frac{\lambda}{\rho}\right)
\label{eq:1.8}
\eeq
and the canonical 1-form
\beq
\int \rmd r\, \theta\,  \rmd \rho
\label{eq:1.9}
\eeq
leads to the Poisson bracket
\beq
\{ \theta(t, \r), \rho (t, \r') \} = \delta (\r - \r')
\label{eq:1.10}
\eeq
It is straightforwardly verified that Eqs.~(\ref{eq:1.1}) and (\ref{eq:1.6})
are a Hamiltonian system with the above bracket.

This model being nonrelativistic possesses the appropriate
nonrelativistic symmetry, namely, Galileo invariance, and
as a consequence of Noether's theorem, there are constants of motion, which
generate via bracketing infinitesimal Galileo transformations.  For future
reference, I record these.  Time translation, space translation, and rotation
act on the coordinates $(t,
\r)$ in the obvious fashion and the transformed fields
$(\rho,\theta)$ are evaluated on the transformed coordinates.  The
corresponding constants of motion are energy $E$, momentum $\bP$, and angular
momentum $L^{ij}$, given by the formulas
\begin{eqnarray}
\hbox{energy: }   E = H =\int \rmd r\, \Hh \nonumber \\[1ex]
\bigl( \hbox{energy density: } \Hh  =  \rho (\boldnab \theta)^2/2m +
\lambda/\rho \bigr)
\label{eq:1.11} \\[1ex]
\hbox{momentum: }  \bP  = 
\int \rmd r\, \bcP \nonumber \\[1ex]
\bigl( \hbox{momentum density: } \bcP  =  \rho \boldnab \theta  \bigr)
\label{eq:1.12} \\[1ex]
\hbox{angular momentum:\qquad\qquad\qquad}\nonumber\\
  L^{ij}  =  \int \rmd r  (r^i \Pp^{\!\! j}
- r^j \Pp^{\!\! i})
\label{eq:1.13} 
\end{eqnarray}
Additionally  there are the Galileo boosts, which boost the spatial
coordinate by a velocity $\bu$
\beq
\r \to \R \equiv \r - t \bu
\label{eq:1.14}
\eeq
While the density field transforms simply
\beq
\rho (t, \r) \to \rho_{\bu} (t, \r) = \rho (t, \R)
\label{eq:1.15a}
\eeq
the velocity potential undergoes an affine transformation,
\beq
\theta (t, \r) \to \theta_{\bu} (t, \r) \equiv \theta (t, \R) +
m(\bu \cdot \r - u^2 t/2)\ \
\label{eq:1.15b}
\eeq
which has the consequence that the velocity acquires (as expected) a
boost
\beq
\bv (t, \r) \to \bv_{\bu} (t, \r) = \bv (t, \R) + \bu
\label{eq:1.15c}
\eeq
The associated constant of motion is the boost generator
\beq
\qquad\hbox{boost generator:}\quad  \B = t \bP - m \int\! \rmd r\, \r \rho
\label{eq:1.16}
\eeq
Also matter is conserved, as a consequence of invariance against a shift of
$\theta$ by constant
\begin{eqnarray}
\theta & \to & \theta + {\rm constant}
\label{eq:1.17a} \\[1ex]
\rho & \to & \rho
\label{eq:1.17b}
\end{eqnarray}
with associated constant of motion
\beq
\qquad\hbox{total matter:} \quad N=\int \rmd r\, \rho
\label{eq:1.18}
\eeq

One can verify that the action $I_{NR}= \int \rmd t\, L_{NR}$ is invariant
against all these transformations and consequently a transformation of a
solution to the equations of motions (\ref{eq:1.1}) and (\ref{eq:1.6})
 is again a solution.  The generators, which can be obtained from the
Lagrangian (\ref{eq:1.7}) by Noether's theorem, are all time-independent,
as can be verified by differentiating them with respect to time, and
evaluating
$(\dot{\rho}, \dot{\theta})$ from the equations of motion
(\ref{eq:1.1}) and (\ref{eq:1.6}).

Note there is a total of $\half (d+1)(d+2)+1$ generators, the correct number
for the (centrally extended) Galileo group, where the extension 1-cocycle and
2-cocycle are responsible for the inhomogeneous term in (\ref{eq:1.15b}) and
lead to the generator $N$ of (\ref{eq:1.18}).

The remarkable fact about the Chaplygin gas is that (in any number of
dimensions) it possesses further symmetries.  First of all one can rescale
time
\beq
t \to T = e^\omega t
\label{eq:1.19}
\eeq
The fields undergo an additional and opposite rescaling
\begin{eqnarray}
\theta (t, \r) & \to & \theta_\omega (t, \r) = e^\omega \theta (T, \r)
\label{eq:1.20a} \\[1ex]
\rho (t, \r) & \to & \rho_\omega (t, \r) = e^{-\omega} \rho (T, \r)
\label{eq:1.20b}
\end{eqnarray}
The time-independent generator reads
\beq
D=tH -\int \rmd r \rho \theta
\label{eq:1.21}
\eeq
Furthermore, a peculiar field-dependent diffeomorphism, which mixes
independent variables $(t, \r)$ and dependent fields $(\rho, \theta)$ also
leaves the action invariant.  The transformation is parameterized by a
d-component vector $\bW$.  On coordinates this acts as 
\begin{eqnarray}
t & \to & T=t+\bW \cdot \r + \half \omega^2 \theta (T, \R)/m
\label{eq:1.22a} \\[1ex]
\r & \to & \R = \r + \bW \theta (T, \R)/m
\label{eq:1.22b}
\end{eqnarray}
Fields transform according to
\begin{eqnarray}
\theta (t, \r) & \to & \theta_{\bW} (t, \r) = \theta (T, \R)
\label{eq:1.23a} \\[1ex]
\rho (t, \r) & \to & \rho_{\bW} (t, \r) = \rho (T, \R) \frac{1}{|J|}
\label{eq:1.23b}
\end{eqnarray}
Here $|J|$ is the Jacobian of the transformation
\beq
J=\det
\left[
\begin{array}{cc}
\frac{\partial T}{\partial t} & \frac{\partial{\bf R}}{\partial t} \\[1ex]
\frac{\partial T}{\partial \r} & \frac{\partial \R}{\partial \r} 
\end{array}
\right]
\label{eq:1.24}
\eeq
The vectorial, $d$-component generator is 
\beq
\G = \int \rmd r \{ \r \Hh - \theta \bcP/m \}
\label{eq:1.25}
\eeq
Just as with the conventional transformations/symmetries, the above
transformations leave the action invariant, and thus take solutions into new
solutions; the additional generators $(D,\G)$, $d+1$ in number, are gotten by
Noether's theorem and are time-independent by virtue of the equations of
motion.

Using the canonical commutator (\ref{eq:1.10}) and the explicit formulas for the
generators, one may compute their Lie algebra.  As is expected, the
$\half(d+1)(d+2)+1$ Galileo generators (\ref{eq:1.11})--(\ref{eq:1.13}),
(\ref{eq:1.16}) and (\ref{eq:1.18}), close on the (extended) Galileo algebra, in
$(d,1)$ space-time.  Supplementing these with the $(d+1)$ additional generators
(\ref{eq:1.21}) and (\ref{eq:1.25}) one arrives at a total of
$\half(d+2)(d+3)$ generators, and their algebra closes on the Poincar\'e
group in one dimension higher,  namely, $(d+1,1)$ space-time.  Moreover,
one establishes that the quantities $(t, \theta, \r)$ transform as light-cone
components of a $(d+2)$ Lorentz vector, with $t$ acting as the $+$ component,
and $\theta$ as the $-$ component~\cite{ref:6}.

Thus we conclude that the nonrelativistic, Galileo invariant Chaplygin gas in
$(d,1)$ space-time possesses a hidden dynamical Poincar\'e symmetry
appropriate to
$(d+1,1)$ space-time, which is realized nonlinearly with field-dependent
diffeomorphisms.

Using symmetry one can generate new solutions from old ones.  Of course when
the transformations are of the familiar Galileo form, the ``new'' solutions
bear an obvious relation to the old ones; they are time or space translated,
space rotated or boosted, or $\theta$-shifted.  However, when the
transformations belong to the hidden symmetry, the new solutions take a new
and unexpected form.

\begin{figure}[htb]
\centerline{\BoxedEPSF{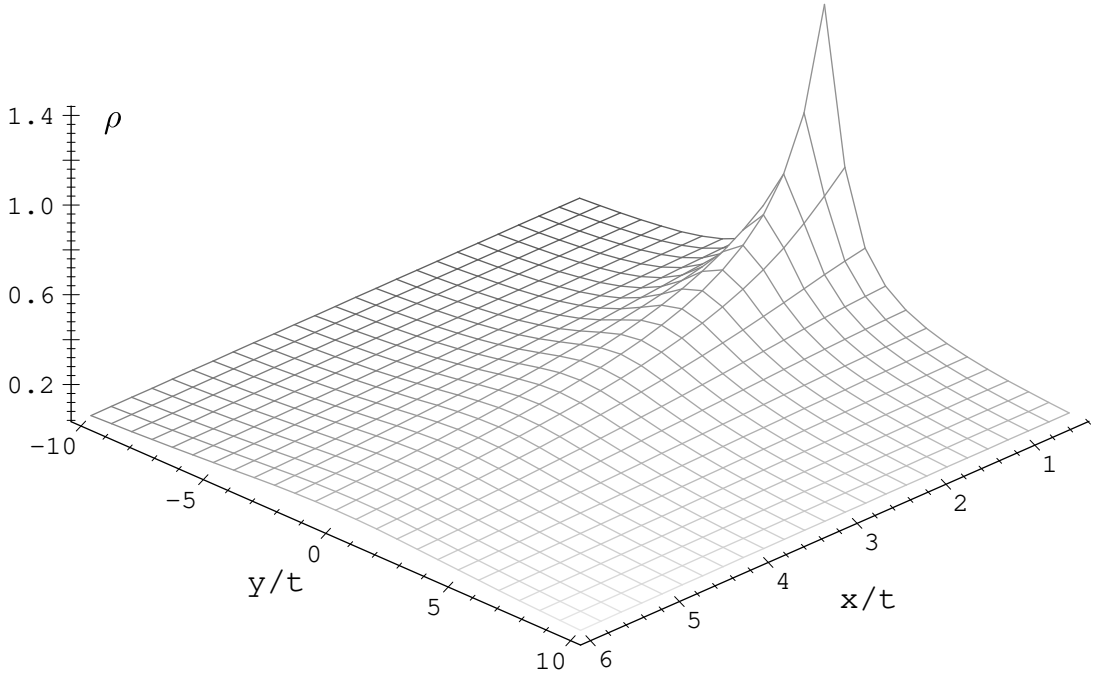 scaled 666}}
\caption{The original density $\rho(t,\r) \propto |t|/r$ (in two spatial
dimensions); $\r=(x,y)$.}\label{fig:1}
\end{figure}

\begin{figure}[htb]
\vspace{9pt}
\centerline{\BoxedEPSF{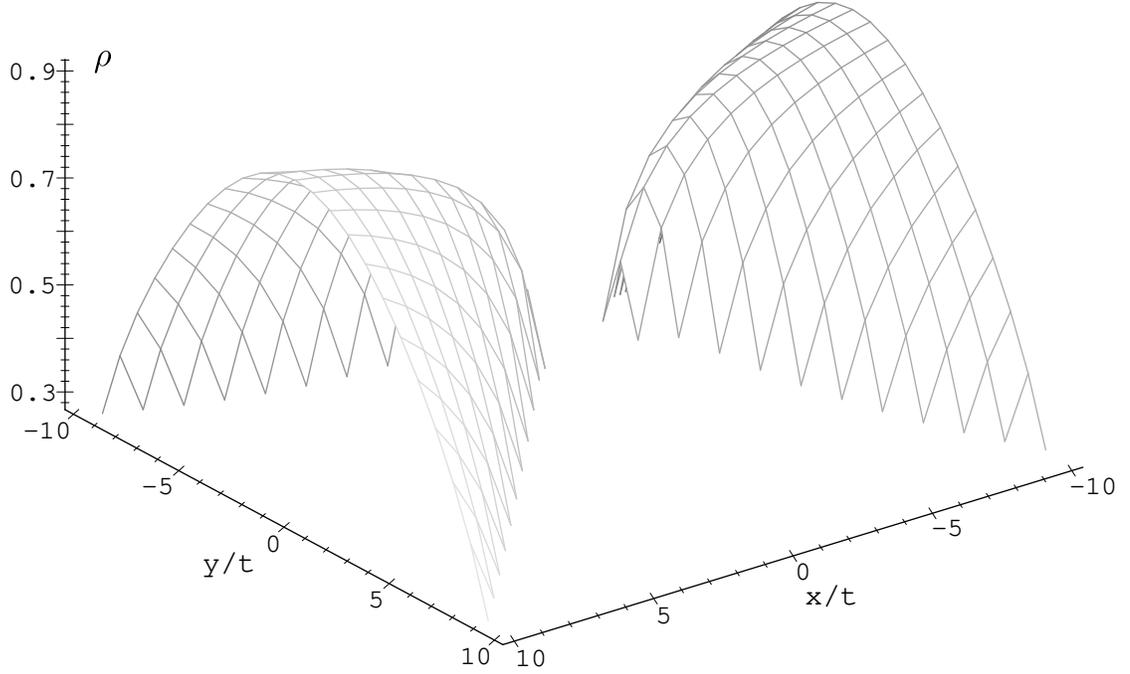 scaled 666}}
\caption{The transformed density $\rho(t,\r)$.}\label{fig:2}
\end{figure}

For example, when $d>1$, a simple solution to (\ref{eq:1.1})
and (\ref{eq:1.6}) is
\beq
(d>1)\colon\ \theta = \frac{-mr^2}{2(d-1)t} , \; 
\rho=\sqrt{\frac{2\lambda}{md}} \, (d-1) \frac{|t|}{r}\ \
\label{eq:1.26}
\eeq
This corresponds to a velocity and current
\beq
(d>1)\colon\    \bv = -\frac{1}{(d-1)} \, \frac{\r}{t} , \; 
\bj =-\epsilon (t) \sqrt{\frac{2\lambda}{md}} \, \hat{r}
\label{eq:1.27}
\eeq
[$\epsilon (t)$ is the step function $t/|t|$.] The $(\theta,\rho)$
profiles in (\ref{eq:1.26}) are invariant against the time-rescaling
transformation (\ref{eq:1.19})--(\ref{eq:1.20b}), but the field-dependent
diffeomorphism (\ref{eq:1.22a})--(\ref{eq:1.23b}) alters the solution
drastically.  The analytic expression for the transformed profile is
uninformative, a plot conveys the situation more clearly.  In Figs.~1 and 2,
the 2-dimensional solutions are plotted:  Fig.~\ref{fig:1} presents
$\rho$ of (\ref{eq:1.26}); in Fig.~\ref{fig:2} the transformed $\rho$ is
exhibited~\cite{ref:7}.

In one spatial dimension, the equations are completely
integrable~\cite{ref:8}.  A hint for this is seen in the special solution,
where the current exhibits a soliton profile.
\begin{eqnarray}
(d=1)\colon\quad \theta &=& -\frac{m}{2k^2 t} \cosh^2 kx \nonumber \\[1ex]
\rho &=&  \sqrt{\frac{2\lambda}{m}} \, \frac{k|t|}{\cosh^2 kx}  \nonumber
\\[1ex]
j &=& -\epsilon (t) \sqrt{\frac{2\lambda}{m}} \tanh kx
\label{eq:1.28}
\end{eqnarray}

To conclude this Section, let me remark that the new symmetries, which we
have uncovered, act equally well on the noninteracting $(\lambda=0)$ model,
for which in fact a complete solution can be given in terms of initial data
for $\rho$ and $\bv$.
\beq
\rho(t, \r) \big|_{t=0}= \rho_0 (\r) \qquad
\bv(t, \r) \big|_{t=0} =\bv_0 (\r)
\label{eq:1.29a}
\eeq
Define the ``retarded'' position of $\q(t,
\r)$ by
\beq
\q + t \bv_0 (\q) = \r
\label{eq:1.29b}
\eeq
and the solution to (\ref{eq:1.1}), (\ref{eq:1.2}) without interaction reads
\begin{eqnarray}
\bv(t, \r)  &=& \bv_0 (\q) \\[1ex]
\rho (t, \r)  &=& \rho_0 (\q) \Bigl| \det \frac{\partial q^j}{\partial r^j}
\Bigr| 
\label{eq:1.29c}
\end{eqnarray}

When $\lambda \ne 0$, one can eliminate $\rho$ in favor of $\theta$ and
present a Lagrangian equivalent to (\ref{eq:1.7}) as
\beq
L_\lambda = -2\lambda \int \rmd r \sqrt{\dot{\theta} + (\boldnab \theta)^2/2m}
\label{eq:1.30}
\eeq
Note that the equations of motion for $\theta$ no longer involve $\lambda$,
which serves merely to normalize the Lagrangian.  In spite of its peculiar
appearance (\ref{eq:1.30}) defines a Galileo invariant theory, which also
possesses the additional symmetries (\ref{eq:1.20a}) and (\ref{eq:1.23a}).

\section{Relativistic Model}

It is possible to give a relativistic generalization of the Chaplygin gas. 
The appropriate Lagrangian in d-spatial dimensions is
\begin{eqnarray}
L_R  &=& \!\int\! \rmd r\, \big[\theta \dot{\rho} -\big(
\sqrt{\rho^2c^2+a^2}
\, 
\sqrt{m^2c^2 + (\boldnab \theta)^2}\, \big)\big] \nonumber \\[1ex]
  &=&\int \rmd r\,  [\theta \dot{\rho} - \Hh_{\!\! R}] 
\label{eq:2.1}
\end{eqnarray}
Here $a$ is a measure of the interaction.  When $a=0$, the above is a
relativistic generalization (\ref{eq:1.7}) with $\lambda=0$.  Retaining $a$,
and letting $c\to \infty$ one finds that
\beq
L_R = - Nmc^2 + L_{NR} \bigr|_{\lambda=ma^2/2}
\label{eq:2.2}
\eeq
Although not manifestly so, $L_R$ is Lorentz and Poincar\'e invariant in
$(d,1)$ space-time (see below), and also matter conservation is respected. 
Thus there are $\half (d+1)(d+2)+1$ generators, where the first number counts
the Poincar\'e generators and ``$+1$'' refers to $N$.

When $a$ vanishes, the model is free and elementary; a complete solution
exists.  Indeed the equations of motion take the form (\ref{eq:1.1}) and
(\ref{eq:1.2}) (with $\f$ set to zero) but $\bv$ is not $\boldnab \theta/m$,
rather it is $c\boldnab \theta/\sqrt{m^2c^2 +(\boldnab \theta)^2}\,$, so that
$\boldnab \theta/m=\bv/\sqrt{1-v^2/c^2}$.  Consequently the solutions take
the same form as in (\ref{eq:1.29a})--(\ref{eq:1.29c}).

Just as in the nonrelativistic case, when $a \ne 0$, $\rho$ can be
eliminated, leaving the  Lagrangian of a
``Born-Infeld''-type model.
\beq
L_a = -a \int \rmd r \sqrt{m^2c^2 - (\partial_\mu \theta)^2}
\label{eq:2.3}
\eeq
And again the coupling strength disappears from the equation of motion for
$\theta$, serving merely to normalize the Lagrangian.  Manifest Poincar\'e
symmetry is now evident, and its generators can be
constructed in the usual fashion from the energy momentum tensor for
(\ref{eq:2.3}).

In view of its analogies to the nonrelativistic model, there is no
surprise that the relativistic model also admits additional hidden symmetries
which leave the action invariant and take solutions into new solutions.  Once
can reparameterize time through a field-dependent transformation, which
depends on the scalar $\omega$
\begin{eqnarray}
t \to T&& \nonumber\\
T &=& \frac{t}{\cosh mc^2 \omega} + \frac{\theta (T,
\r)}{mc^2} \tanh mc^2 \omega 
\label{eq:2.4} \\[1ex]
\theta(t, \r)  & \to& \theta_\omega (t, \r) \nonumber\\
\theta_\omega (t, \r)  &=&
\frac{\theta (T,\r)}{\cosh mc^2
\omega} -mc^2\tanh mc^2 \omega 
\label{eq:2.5}
\end{eqnarray}
The associated conserved  generator is
\beq
D =  \int \rmd r\, (m^2c^4 t \rho +\theta\Hh_{\!\! R})
\label{eq:2.6}
\eeq
Also there is a spatial reparameterization, governed by the
d-component vector $\bW$ ($\omega \equiv |\bW|$).
\begin{eqnarray}
\r \to \R &&\nonumber\\
\R &=&   \r-\bW \theta(t, \R) 
\frac{\tan mc\omega}{mc\omega} \nonumber\\
 &&\quad{} +
\bW \bW \cdot \r \frac{1-\cos mc\omega}{\omega^2 \cos mc\omega}
\label{eq:2.7} \\[1ex]
\theta(t, \r) &\to& \theta_{\bW} (t, \r)\nonumber\\
\theta_{\bW} (t, \r) &=&   \frac{\theta (t,\R)}{\cos mc
\omega} -mc \bW \cdot \r \frac{\tan mc \omega}{\omega} 
\label{eq:2.8}
\end{eqnarray}
with conserved vectorial generator
\beq
\G =  \int \rmd r (m^2c^2 r\rho +\theta \bcP)
\label{eq:2.9}
\eeq
Only the $\theta$ transformation law is exhibited; the one for $\rho$ can be
deduced from the equation of motion for $\theta$, which follows from
(\ref{eq:2.1}).

The additional symmetries give us $d+1$ further generators, which
supplement the previously described $\half (d+1)(d+2)+1$ generators, for a
total of
$\half (d+2)(d+3)$ -- just the right number for the Poincar\'e group in
$(d+1, 1)$ space-time.  And indeed upon computing the canonical Lie algebra
brackets of all the generators one finds that the totality of generators
closes on the
$(d+1,1)$ Poincar\'e group.  The computation is based on the same bracket
(\ref{eq:1.10}) as in the nonrelativistic case [because the canonical 1-form
of (\ref{eq:2.1}) coincides with that of (\ref{eq:1.7}), and is given in
(\ref{eq:1.9})].  Moreover the set of quantities $(t, \theta, \r)$ transforms
as a $(d+2)$-Lorentz vector in Cartesian components~\cite{ref:6}.

As in the nonrelativistic case, one may use the additional hidden symmetry
transformations to map solutions into new solutions with different
properties.  Additionally, one may use the relativistic--nonrelativistic
connection to obtain solutions of the Chaplygin gas problem by taking the
$c\to \infty$ limit of the Born-Infeld solutions~\cite{ref:8a}.  The Born-Infeld
model in one spatial dimension is completely integrable~\cite{ref:9}.

\section{Common Ancestry}

The ``hidden'' symmetries and the associated haphazard transformation laws
may be given a coherent setting by considering the Nambu-Goto action for a
d-brane in $(d+1)$ spatial dimensions, moving on $(d+1,1)$ dimensional
space-time.  [A~d-brane is a d-dimensional extended object: 1-brane is a
string, 2-brane is a membrane, and so on.  A~d-brane in $(d+1)$ space
divides that space in two.]

The Nambu-Goto action reads
\begin{eqnarray}
I_{\rm NG} &=& \int \rmd \phi^0 \rmd \phi^1 \cdots \rmd \phi^d \sqrt{G}
\label{eq:3.1} \\[1ex]
G &=& (-1)^d \det \frac{\partial X^\mu}{\partial \phi^\alpha}
\frac{\partial X_\mu}{\partial \phi^\beta}
\label{eq:3.2}
\end{eqnarray}
Here $X^\mu$ is a $(d+1,1)$ target space-time (d-brane) variable, with $\mu$
extending over the range $\mu=0,1,\dots,d,d+1$.  The $\phi^\alpha$ are
variables describing the extended object with $\alpha$ ranging
$\alpha=0,1,\dots,d$; $\phi^\alpha$,
$\alpha=1, \dots, d$, parameterizes the d-dimensional d-brane, while the
extended object evolves in $\phi^0$.

The Nambu-Goto action is parameterization invariant, and we shall show that
two different parameterizations (``light-come'' and ``Cartesian'') result in
the Chaplygin gas and Born-Infeld actions.  The parameterizations are fixed as
follows.  For both parameterizations we choose $(X^1, \dots, X^d)$ to
coincide with $(\phi^1,
\dots, \phi^d)$ and rename them $\r$ (a d-dimensional vector).

For the light cone parameterization we define $X^\pm$ as $\frac{1}{\sqrt{2}}
(X^0 \pm X^{d+1})$.  $X^+$ is renamed $t$ and identified with $\sqrt{2\lambda
m}\, \phi^0$.  This completes the fixing of the parameterization and the
remaining variable is $X^-$, which is a function of $\phi^0$ and $\boldphi$,
or after redefinitions, of $t$ and $\r$.  $X^-$ is renamed as
 $\theta(t,\r)/m$ and then the Nambu-Goto action (\ref{eq:3.1}) in this
parameterization coincides with the Chaplygin gas action $\int \rmd t
L_\lambda$, where
$L_\lambda$ is given in (\ref{eq:1.30}).

For the second, Cartesian parameterization $X^0$ is renamed $ct$ and
identified with $amc \phi^0$.  The remaining target space variable $X^{d+1}$,
a function of $\phi^0$ and $\boldphi$, equivalently of $t$ and $\r$, is
renamed
$\theta(t,\r)/mc$.  Then the Nambu-Goto action (\ref{eq:3.1}) reduces
to the Born-Infeld action $\int \rmd t L_a$, (\ref{eq:2.3}).

The relation to the Nambu-Goto action explains the origin of the hidden
$(d+1,1)$ Poincar\'e group in our two nonlinear models on $(d,1)$ space-time:
Poincar\'e invariance is what remains of the reparameterization invariance of
the Nambu-Goto action after choosing either the light-cone or Cartesian
parameterizations.  Also the nonlinear, field dependent form of the
transformation laws (\ref{eq:1.19})--(\ref{eq:1.20b}),
(\ref{eq:1.22a})--(\ref{eq:1.23b}), (\ref{eq:2.4}), (\ref{eq:2.5}),
(\ref{eq:2.7}), (\ref{eq:2.8}) is understood: it arises from the identification
of some of the dependent variables ($X^\mu$) with the independent variables
$(\phi^\alpha)$.

The complete integrability of the $d=1$ Chaplygin gas and Born-Infeld model
is a consequence of the fact that both descend from a string in 2-space.  But
the Nambu-Goto theory for that system is completely
integrable~\cite{ref:6}.

We observe that in addition to the nonrelativistic descent from the
Born-Infeld theory to the Chaplygin gas, there exists a mapping of one system
on another, and between solutions of one system and the other, because both
have the same d-brane ancestor.  The mapping is achieved by passing from the
light-cone parameterization to the Cartesian, or vice-versa~\cite{ref:8a}. 
Specifically this is accomplished as follows: 

\paragraph{{\rm Chaplygin gas $\to$ Born-Infeld:}}  
Given $\theta_{NR} (t,\r)$, a
nonrelativistic solution, determine $T(t,\r)$ from the equation
\beq
T+\frac{1}{mc^2} \theta_{NR} (T,\r) = \sqrt{2} \, t
\label{eq:3.3}
\eeq
Then the relativistic solution is
\beq
\theta_R(t,\r)=\frac{1}{\sqrt{2}} mc^2 T - \frac{1}{\sqrt{2}}
\theta_{NR} (T,\r)
\label{eq:3.4}
\eeq

\paragraph{{\rm Born-Infeld $\to$ Chaplygin gas:}}
Given $\theta_R(t,\r)$, a relativistic solution, find $T(t,\r)$ from
\beq
T+\frac{1}{mc^2} \theta_{R} (T,\r) = \sqrt{2} \, t
\label{eq:3.5}
\eeq
Then the nonrelativistic solution is
\beq
\theta_{NR}(t,\r)=\frac{1}{\sqrt{2}} mc^2 T - \frac{1}{\sqrt{2}}
\theta_{R} (T,\r)
\label{eq:3.6}
\eeq
All the relationships are summarized in Fig.~3.
\begin{figure}[htb]
\centerline{\BoxedEPSF{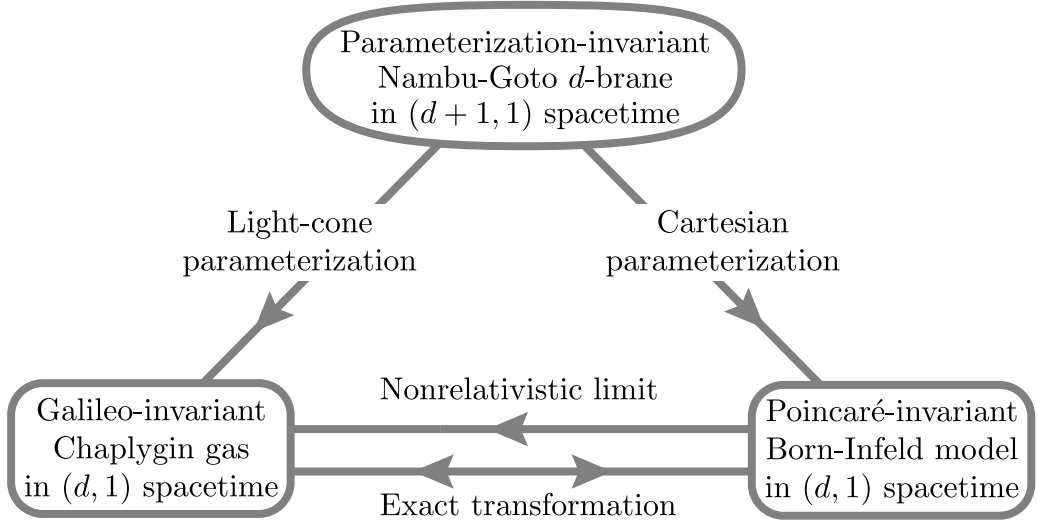}}
\caption{Dualities and other relations between nonlinear
equations.}\label{fig:3}
\end{figure}

One cannot establish the connection of our two nonlinear equations to the
Nambu-Goto action in absence of the interaction, neither in the nonrelativistic
$(\lambda=0)$ nor relativistic
$(a=0)$ cases: one cannot eliminate $\rho$ in favor of $\theta$, because in
the absence of an interaction
$\rho$ no longer appears in the equation of motion for $\theta$. 
Equivalently one sees that the $\theta$-Lagrangians $L_\lambda$
(\ref{eq:1.30}) and $L_a$ (\ref{eq:2.3}) vanish with $\lambda$ and $a$,
respectively.  The Nambu-Goto action is normalized by the d-brane tension,
which has been scaled to unity in (\ref{eq:3.1}).  Thus the nonrelativistic
and relativistic free models $(\lambda=0=a)$ in their $\rho-\theta$ forms,
(\ref{eq:1.7}) and (\ref{eq:2.1}) respectively, may be viewed as
a parameterized description of ``tension-less'' d-branes.  

Finally we remark
that the emergence of the $(d+1,1)$ Poincar\'e group from the $(d,1)$ Galileo
group can also be understood in Kaluza-Klein-like construction~\cite{ref:10}.



\begin{thebibliography}{99}
\bibitem{ref:7} Analytic expressions for the
transformed solutions are given in the first paper of Ref.~protect\cite{ref:1}.
\bibitem{ref:8a} Examples are worked out in \protect\cite{ref:2}.
\end{thebibliography}
\end{document}